\documentclass[revtex, amsmath, amssymb,superscriptaddress, reprint]{revtex4-2}

\usepackage[utf8]{inputenc}
\usepackage{graphicx}
\usepackage{dcolumn}
\usepackage{bm}
\usepackage[utf8]{inputenc}
\usepackage[T1]{fontenc}
\usepackage{mathptmx}
\usepackage{gensymb}
\usepackage{upgreek}
\usepackage{xcolor}
\usepackage{textcomp}
\usepackage{hyperref}
\usepackage{mhchem}
\usepackage{lineno}
\hypersetup{colorlinks=true,linkcolor=red,citecolor=blue,filecolor=magenta,urlcolor=magenta}

\usepackage{titlesec}
\titleformat{\section}[hang]{\small\bfseries\sffamily}{\thesection.}{0.5em}{\MakeUppercase}
\titlespacing{\section}{0pc}{1pc}{0.2pc}

\begin{document}

\title{SmartCut \ce{Er}:\ce{LiNbO3} with high optical coherence enabling optical thickness control
}


\author{Sihao Wang}
\affiliation{Department of Electrical Engineering, Yale University, New Haven, CT 06511, USA}
\author{Likai Yang}
\affiliation{Department of Electrical Engineering, Yale University, New Haven, CT 06511, USA}
\author{Mohan Shen}
\affiliation{Department of Electrical Engineering, Yale University, New Haven, CT 06511, USA}
\author{Wei Fu}
\affiliation{Department of Electrical Engineering, Yale University, New Haven, CT 06511, USA}
\author{Yuntao Xu}
\affiliation{Department of Electrical Engineering, Yale University, New Haven, CT 06511, USA}
\author{Rufus L. Cone}
\affiliation{Department of Physics, Montana State University, Bozeman, MT 59717, USA}
\author{Charles W. Thiel}
\affiliation{Department of Physics, Montana State University, Bozeman, MT 59717, USA}
\author{Hong X. Tang}
\email{hong.tang@yale.edu}
\affiliation{Department of Electrical Engineering, Yale University, New Haven, CT 06511, USA}


\begin{abstract}
Integrated photonics capable of incorporating rare earth ions with high optical coherence is desirable for realizing efficient quantum transducers, compact quantum memories, and hybrid quantum systems. Here we describe a photonic platform based on the SmartCut erbium-doped lithium niobate thin film, and explore its stable optical transitions at telecom wavelength in a dilution refrigerator. Optical coherence time of up to 180\,$\mu$s, rivaling the value of bulk crystals, is achieved in optical ridge waveguides and ring resonators. With this integrated platform, we demonstrate tunable light-ion interaction and flexible control of optical thickness by exploiting long waveguides, whose lengths are in principle variable. This unique ability to obtain high optical density using a low concentration ions further leads to the observation of multi-echo pulse trains in centimeter-long waveguides. Our results establish a promising photonic platform for quantum information processing with rare earth ions.

\end{abstract}
\maketitle

\section{Introduction}
Atomic and atomic-like defects in solid crystals provide a robust avenue for light-ion interactions and have been promising candidates for a variety of quantum information processing technologies \cite{togan2010quantum, de2012quantum, gao2012observation, taminiau2014universal, bradley2019ten, raha2020optical}. Rare earth ions (REIs) are among the popular choices of study \cite{wesenberg2007scalable, simon2010quantum, tittel2010photon} thanks to many favorable optical and spin properties. The stable optical transitions, long population relaxation time and high fluorescence quantum efficiencies make them ideal for nonlinear optics applications \cite{thiel2011rare}. The narrow homogeneous linewidth \cite{equall1994ultraslow, bottger2009effects} allows encoding quantum information in the narrowly burnt spectral holes \cite{babbitt2014spectral}. The weak interaction of the 4f electrons with the host environment \cite{macfarlane1987coherent} not only allows them to exhibit long coherent spin states \cite{zhong2015optically}, but also relaxes the requirement for the host crystals. With commonly used host materials like yttrium orthosilicate (Y$_2$SiO$_5$), yttrium orthovanadate (YVO$_4$), and lithium niobate (LiNbO$_3$), it is advantageous to leverage this flexibility of choices to tailor for particular applications.

 By directly doping REIs in the crystal growth stage, for example within a melt, uniform ion distribution at desired sites can be well-achieved in bulk hosts, granting high material quality. Integrated optical platforms, on the other hand, enable low-loss waveguides and high quality factor micro-resonators with small mode profile, making them appealing in REI applications to enhance light-ion coupling. This raises the interest of integrating well-studied bulk REI doped crystals with on-chip photonics. Evanescent coupling of REIs to optical nanostructures has been realized by placing doped crystals in close proximity via bonding or waveguide deposition technique \cite{yang2021photonic,dibos2018atomic,miyazono2017coupling}. Resonators with quality factor up to millions are readily achievable, but the small mode overlap makes the coupling less efficient. Alternatively, direct ion milling of the host crystal to fabricate optical cavities has been demonstrated. While maintaining complete mode overlap, the quality of optical cavities is still low (Q $\sim$ 4000) \cite{zhong2017nanophotonic, zhong2017interfacing, craiciu2019nanophotonic}. Therefore, it is imperative to find a suitable platform that allows full mode overlap and versatile fabrication of high quality optical structures.

In this work, we exploit a promising platform based on the SmartCut thin films prepared from bulk erbium doped lithium niobate (Er:LN). Thin film LN has garnered a vast interest in making efficient, compact and high-performance nonlinear integrated photonic circuits since its breakthrough in the thin-film nanofabrication technology \cite{zhang2017monolithic}. Compared to direct implantation or in-diffusion of REIs into the thin film \cite{wang2020incorporation}, the SmartCut technique \cite{hu2012lithium, poberaj2012lithium, bazzan2015optical} avoids damage from the doping process and the necessity of high-temperature annealing, which can potentially improve material quality for better coherence and greatly facilitates the fabrication process. The ion distribution in SmartCut thin films is also more uniform, providing a full mode overlap. 
While several studies \cite{dutta2019integrated, chen2021efficient, xu2021er3} have investigated the optical properties of REIs doped LN thin films, most focus on classical applications and a thorough characterization of the coherence properties is still lacking. This is essential if REI doped thin films are to be used for quantum applications. Furthermore, efficient absorption of input photons in a compact device without compromising coherence properties is desirable in optical quantum storage and processing \cite{tittel2010photon, mazelanik2019coherent}. This presents opportunities for high-density integrated quantum photonic circuits. Here, we present our study of SmartCut Er:LN thin films. Low-loss ring resonators (Q $\sim$ 1 million) and centimeter-long waveguides are fabricated and characterized in a dilution refrigerator. Photon echo measurements are performed to probe coherence under various temperatures and external magnetic fields, yielding comparable results with bulk crystals. A lengthened coherence time of up to 180 $\mu$s is achieved at extremely low temperature of 20 mK and 0.55 T magnetic field. Tailorable optical density in the long waveguide allows the observation of rich phenomena in a compact device through the multi-echo train, showcasing the versatility of the integrated platform.

\section{Experimental Approach}

\begin{figure*}[!htbp]
\includegraphics[width=0.9\textwidth]{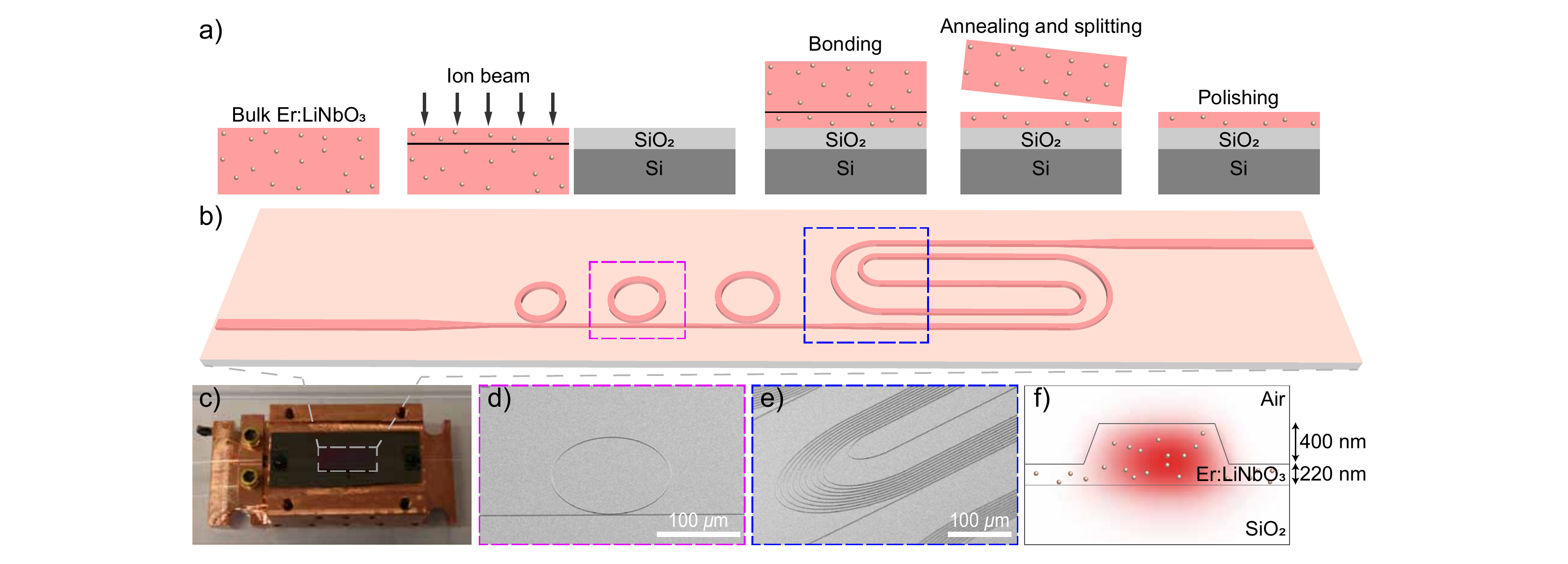}
\caption{\label{fig:1} \textbf{a)} Schematic of the SmartCut process for bonding the erbium doped lithium niobate thin film onto the oxided silicon substrate. \textbf{b)} Illustration of a photonic device with integrated microrings and a long spiral waveguide. \textbf{c)} Image of a packaged device. Optical fibers are glued on the device and anchored on the substrate. The whole device is packaged in a copper box. \textbf{d,e)} SEM images of a ring resonator and a spiral waveguide of 4 cm long. \textbf{f)} Schematic cross section of the waveguide and the overlaid fundamental TE mode profile.}
\end{figure*}

\subsection{Device preparation}
The SmartCut bonding process is illustrated in Fig.~\ref{fig:1}a). Commercial bulk wafers (SurfaceNet) of 100 ppm z-cut Er:LN are chosen for the bonding process (NanoLN) on insulator. During the bonding, helium ions are implanted to form an amorphous layer with a depth depending on the implantation energy. Another thermal oxide wafer is prepared as the substrate. The implanted wafer is subsequently bonded on the thermal oxide substrate. Low temperature annealing of $<$ 200 \textcelsius  \cite{hu2012lithium, poberaj2012lithium} can be used to improve bonding strength. A further increase of annealing temperature ($\sim230$ \textcelsius) causes a splitting at the implantation depth. A final chemical mechanical polishing (CMP) process is used to achieve a flat surface with low roughness. 

Fig.~\ref{fig:1}b) shows the not-to-scale schematic drawing of the devices consisting of multiple ring resonators and a long waveguide. The ring width is 1.8 $\mu$m and ring radii ranging from 65\,$\mu$m to 75\,$\mu$m are chosen to ease the coupling conditions to the bus waveguide. At the intersection with the rings, the width of bus waveguides is tapered to 0.8 $\mu$m for a strong evanescent coupling. The width of the centimeter-long waveguide is 2 $\mu$m. The minimal bending radius is 50 $\mu$m to reduce bending loss. We fabricate these micro-photonic structures by patterning hydrogen silsesquioxane (HSQ) on the Er:LN thin film with electron beam lithography (EBL). The structures are then half etched through reactive ion etching (RIE) with argon plasma. After fabrication, optical fibers are carefully glued to the waveguide facets with a cryogenic glue. The glued device is finally placed in a copper box for protection. Fig.~\ref{fig:1}c) is the photo of the final device. The scanning electron microscope (SEM) images of a typical ring and a long waveguide are shown in Fig.~\ref{fig:1}d,e). The fundamental transverse electric (TE) mode  (Fig.~\ref{fig:1}f)) is well confined in the half-etched waveguide, ensuring a full modal overlap with the Er ions.  

\begin{figure*}[!htbp]
\includegraphics[width=0.9\textwidth]{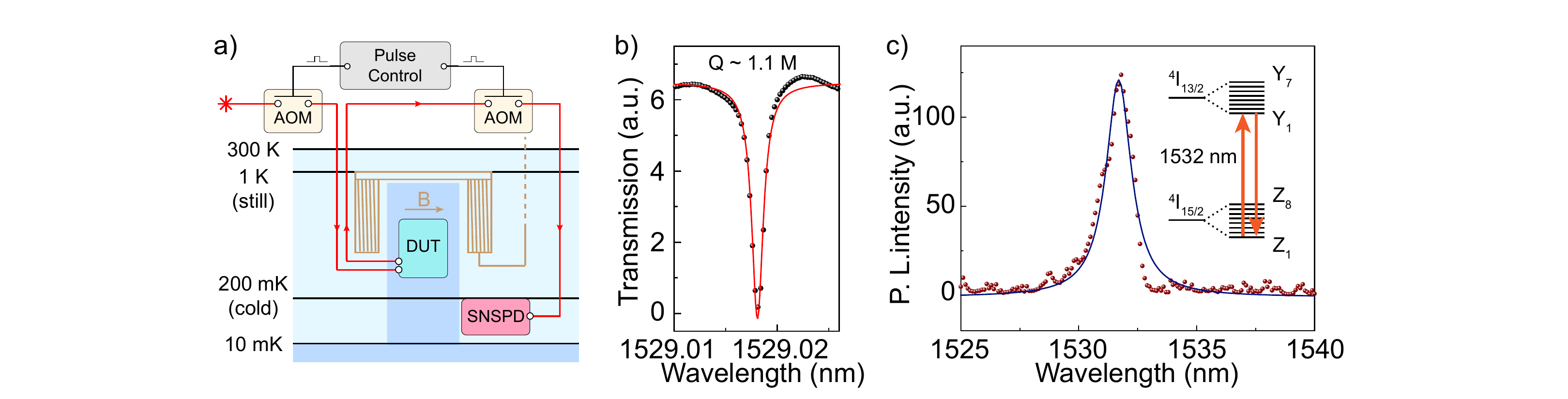}
\caption{\label{fig:2} \textbf{a)} A simplied schematic of the experimental setup. \textbf{b)} A typical resonance of Q $\sim$ 1.1 M at 1529 nm, out of the Er$^{3+}$ inhomogeneous linewidth. \textbf{c)} The inhomogeneous broadening of Er:LN, mapped by the resonant fluorescence. The inset shows the optical transition ($Z_{1}$-$Y_{1}$) between the lowest Kramer doublet states in the ground and the excited energy levels. }
\end{figure*}
\subsection{Experimental setup and device characterization} 
A simplified experimental schematic is shown in Fig.~\ref{fig:2}a). The packaged device is loaded at the mixing chamber of the dilution refrigerator, with a base temperature of 10 mK. Two accousto-optic modulators (AOM) are placed before and after the device for pumping and gating at different time sequences. The output signal is detected by a photodetector (PD) or a fiber-coupled superconducting nanowire single photon detector (SNSPD). A tunable laser is used to excite Er$^{3+}$ ions. Variable attenuators up to 100 dB are inserted after the laser to adjust the input power. A home-made superconduting coil of 20 mT/A is used to provide a uniform magnetic field (B field), parallel to the c-axis of the LN thin film. 

To ensure good film quality and minimal fabrication impact, we measured the Q factors of the ring resonators. The average Q factors is $\sim 800$ k. Multiple resonances with over a million Q are easily observable (Fig.~\ref{fig:2}b)). The activation of Er ions is confirmed through resonant fluorescence measurements, which also maps the inhomogeneous linewidth of the Er ensemble (Fig.~\ref{fig:2}c)). The 166 GHz measured linewidth is comparable to the literature value of 180 GHz in the bulk crystal.

\section{Result}
\begin{figure*}[!htbp]
\includegraphics[width=0.9\textwidth]{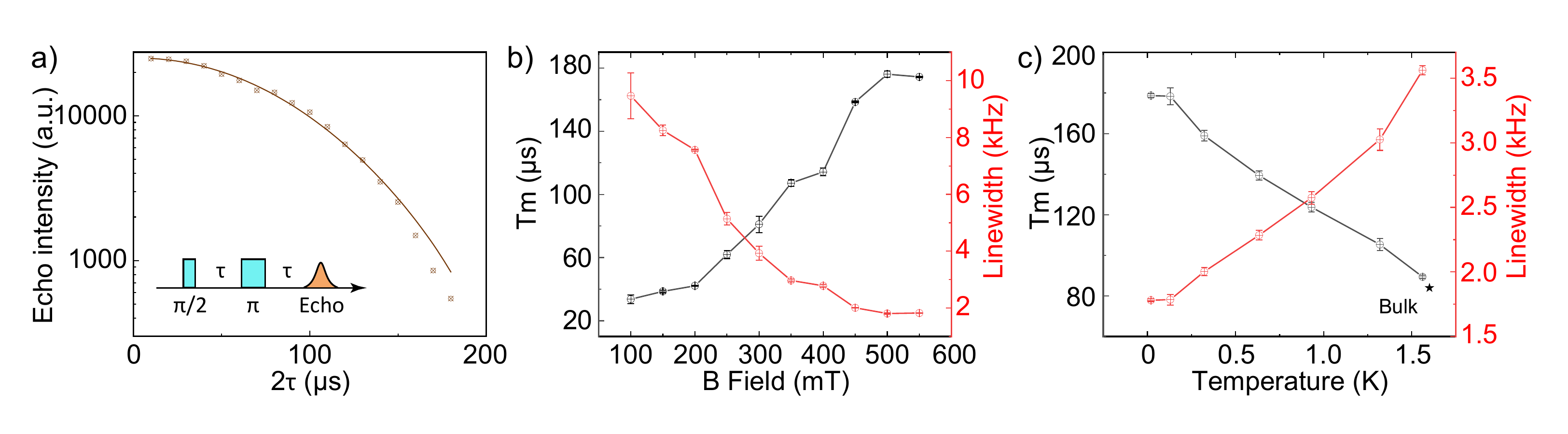}
\caption{\label{fig:3} \textbf{a)} A typical plot of echo area with time delay at 20\,mK under 0.4\,T B field applied parallel to the c-axis of the LN thin film. The inset shows the pulse sequence of a two photon echo scheme. \textbf{b)} The coherence relaxation time and the corresponding effective homogeneous linewidth as a function of B field at 20 mK. \textbf{c)} The coherence relaxation time and the corresponding effective homogeneous linewidth as a function of temperature under 0.5 T of B field. The coherence time of the bulk crystal used for bonding measured at 1.6\,K and 0.5\,T is marked as the star symbol in the plot.}
\end{figure*}
\subsection{Coherence measurement}
The two photon echo technique is commonly used to extract the long phase coherence time as it is \textcolor{red}{less} influenced by the pump laser linewidth. As shown in the inset of Fig.~\ref{fig:3}a), two short optical pulses with a time delay $\tau$ are generated, with duration ranging from 100 ns to 500 ns to optimize the echo signal strength. A B field of up to 0.55 T is applied parallel to the c-axis of lithium niobate to suppress the influence of spectating ions in the crystal lattice. By fitting the area of the echo signal with the total delay 2$\tau$, we are able to extract the phase coherence time  $T_\mathrm{m}$ from the echo intensity $I$($\tau$) \cite{mims1968phase, bottger2006optical},
\begin{eqnarray}
I(\tau)=I_0 \mathrm{Exp} \left[ -2 \left( \frac{2\tau}{T_\mathrm{m}} \right)^x\right ],
\label{eq:3}
\end{eqnarray}
where $x>1$ represents the influence from spectral diffusion. The effective homogeneous linewidth is thus given by $\Gamma_\mathrm{h} = 1/(\pi T_\mathrm{m})$. Fig.~\ref{fig:3}a) shows a typical coherence decay plot at 20 mK temperature and 0.4 T B field. The fitting yields a coherence time of 145 $\mu$s. Fig.~\ref{fig:3}b) shows the variation of the coherence time $T_\mathrm{m}$ with the applied B field strength. The solid lines are connected to guide the viewing of the data. The corresponding plot of the effective homogeneous linewidth is shown in red in the same figure. The measurement was performed at 20 mK, close to the base temperature of the dilution refrigerator. A long coherence time of $T_\mathrm{m} \approx$ 180 $\mu$s ($\Gamma_\mathrm{h} = 1.8$ kHz) is achievable. Fig.~\ref{fig:3}c) shows the variation of the coherence time with temperature, under a constant B field of 0.5 T. It is worth noting that the bulk Er:LN wafer was characterized before sending for SmartCut bonding. The bulk coherence time at 0.5 T and 1.6 K was measured to be 84 $\mu$s, shown as the star symbol in Fig.~\ref{fig:3}c). This is in good agreement with the trend of the data from the Er:LN thin film after the SmartCut bonding, demonstrating that the SmartCut bonding process does not degrade the coherence property of the sample. 

\subsection{Multi-echo train arising from tailored optical thickness}
\begin{figure*}[!htbp]
\includegraphics[width=0.9\textwidth]{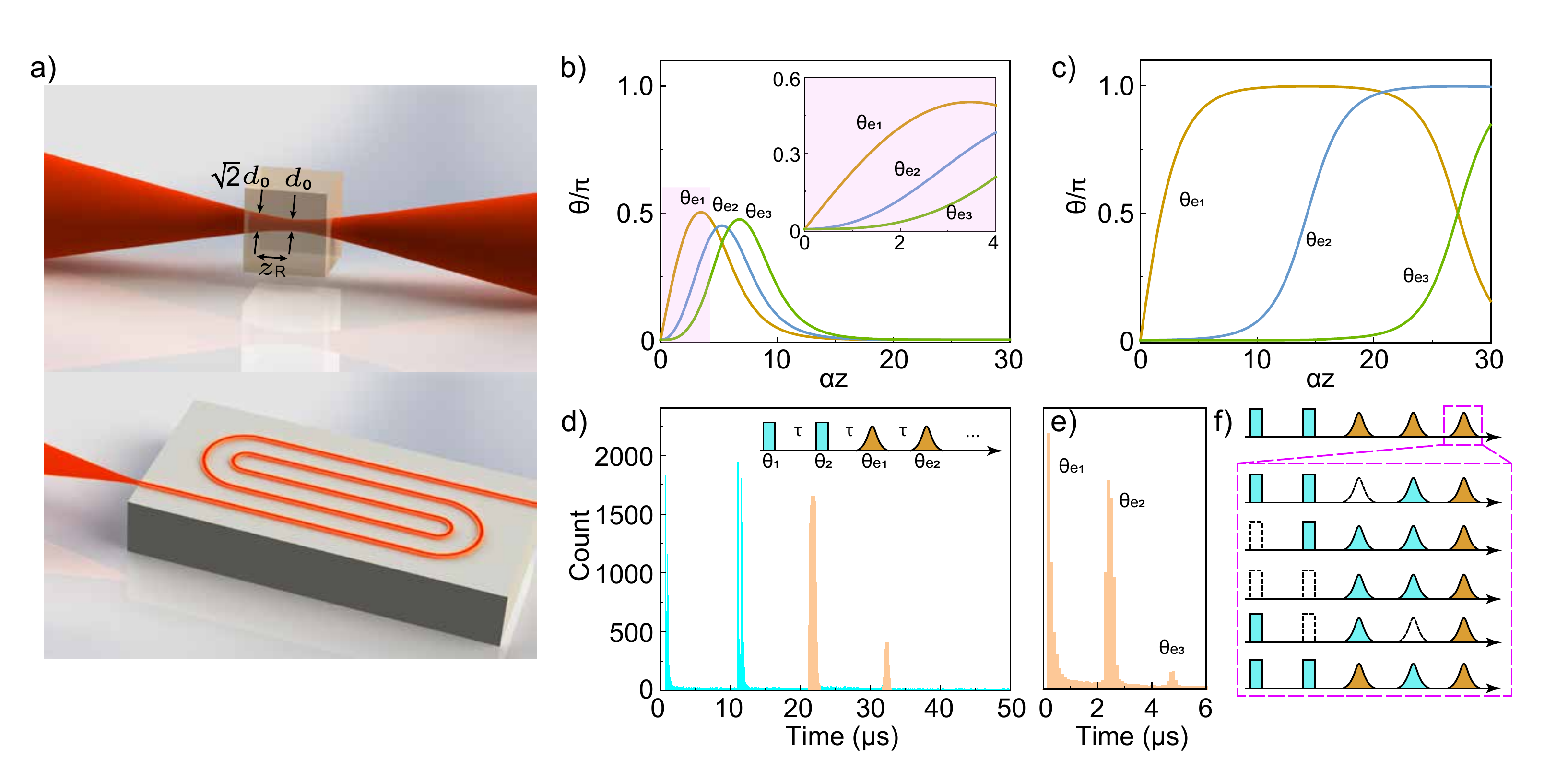}
\caption{\label{fig:4} \textbf{a)} The upper panel is the schematics of free space illumination in a bulk crystal. The effective optical thickness is dependent on the Rayleigh range $z_\mathrm{R}$, defined as the distance from the waist, at which the beam size is doubled. $d_0$ is beam waist diameter. The lower panel shows light inside the waveguide. The effective optical depth is proportional to the waveguide length. \textbf{b)} Numerical analysis of first three echoes at different optical thickness with weak pump pulses $\theta_1 = 0.3\pi$ and $\theta_2 = 0.85\pi$. The inset is the magnified plot near $\alpha z = 2$, showing a nonzero value of all three echo signals. \textbf{c)} Numerical analysis of first three echoes with optimized pump pulses $\theta_1 = 0.499\pi$ and $\theta_2 = 0.999\pi$ in the two photon echo scheme. Each echo saturates at $\pi$ at a sufficiently large optical thickness. \textbf{d)} The measured two-echo sequences. The cyan pulses are from the pump pulses where the orange pulses are echo signals. The third echo signal is too weak to be observed at the current scale. The inset is the schematic of the multi-echo train with constant time delay $\tau$. \textbf{e)} A similar measurement from the same device with a shorter (2\;$\mu$s) delay to demonstrate three echo signals. \textbf{f)} Schematic representations of 5 distinct processes that contribute to the third echo, including self-stimulated echo, two photon echo and primary echo revival. }
\end{figure*}

The optical thickness of a sample is dependent on the material doping concentration and the optical interaction length. Increasing doping concentration generally undermines coherence due to spectral diffusion from long range spin-spin interactions. For a bulk crystal, the optical interaction length depends on the Rayleigh range $z_\mathrm{R}$ of the focusing lens and the sample dimension (Fig.~\ref{fig:4}a), upper panel). Efficient nonlinear interactions between optical photons and ions occur only within the Rayleigh range. An important advantage of Er:LN thin film is the freedom to manipulate the interaction length in long waveguides (Fig.~\ref{fig:4}a), lower panel) while still maintaining a small footprint and good coherence properties from low doping concentration. This allows the observation of rich phenomena in optically dense media and presents the opportunity to achieve a compact and integrated photonic device. Here a 4-centimeter long waveguide is designed to achieve a large optical thickness with a footprint of 2 mm $
\times$ 0.3 mm. For 100 ppm Er:LN at 1532 nm, the absorption coefficient $\alpha$ is 0.7 cm$^{-1}$ for linearly polarized light with E$\perp$z, 0.4 cm$^{-1}$ for B$\perp$z, and 1.1 cm$^{-1}$ when both E$\perp$z and B$\perp$z. Optical thickness of the 4 centimeter long waveguide is estimated to be about $\alpha z \approx 2$. The coherent light-ion coupling at this optical thickness gives rise to new controllable
echo phenomena. As a specific example, we show the observation of the multi-echo train. This phenomenon has been studied in the general framework of the well-known area theorem \cite{mccall1969self, gl1971analytical, eberly1998area, eberly2002wave, chaneliere2014strong, shchedrin2015analytic, gutierrez2016vector, urmancheev2019two, azadeh1998efficient, moiseev2020photon}. In particular, a recent study by S. A. Moiseev, \textit{et al.} \cite{moiseev2020photon} provides a rigorous analytical model to describe the echo pulse trains as a result of the coherent interaction of optical fields with resonating atoms in an optically thick medium. Based on the model, the evolution of the echo area is described by the area theorem,
\begin{eqnarray}
\partial_z \theta = \frac{1}{2} \alpha w_0(z) \sin \theta(z),
\label{eq:4}
\end{eqnarray}
where $\theta = \int_\infty^\infty \Omega(t) dt$ is the pulse area ($\theta_{1,2}$ are the first and the second pump pulses, $\theta_{\mathrm{e1},\mathrm{e2},\mathrm{e3}}$ correspond to the first three echo signals), $\Omega (t)$ is the Rabi frequency of the complex electric field, and $w_0(z)$ is the initial inversion component of the Bloch vector $\left(u(z), v(z), w(z)\right)^\intercal$  describing the atomic system. In the two photon echo scheme, the atomic system is at the initial ground state $w_0 = -1$ before the first pump pulse and modified $w_0 = \cos \theta_1$ before the second pump pulse. The general equation of each individual echo pulse $\theta_{e,n}$ can be obtained with the assumption of non-overlapping pulses ($1/(\pi\Gamma_{\mathrm{inh}}) \ll \Delta t_{1,2} \ll \tau \ll T_\mathrm{m}$, where $\Gamma_{\mathrm{inh}}$ is the inhomogeneous linewidth of the ensemble and $\Delta t_{1,2}$ are the pump pulse durations) \cite{moiseev2020photon},
\begin{eqnarray}
\partial_z \theta_{e,n} = \frac{1}{2} \alpha \left(2 v_0(z) \cos^2 \frac{\theta_{e,n}(z)}{2} + w_0(z) \sin \theta_{e,n}(z) \right),
\label{eq:5}
\end{eqnarray}
where $v_0(z)$ and $w_0(z)$ are the initial values ($t = t_{e,n} - (n+1/2)\tau $) of the out-of-phase and inversion components of the Bloch vector for the $n^{\mathrm{th}}$ echo. 

Numerical analysis of echo area with optical thickness based on Eq.~\ref{eq:5} are shown in Fig.~\ref{fig:4}b) and \ref{fig:4}c). Fig.~\ref{fig:4}b) shows the strength of the first three echoes with respect to different optical thicknesses at weaker pump pulses $\theta_1 = 0.3\pi$ and $\theta_2 = 0.85\pi$. The inset is the magnified plot near $\alpha$z$\approx2$, showing non-zero values of higher order echoes. Fig.~\ref{fig:4}c) shows the same first three echoes with optimized pulse strengths in the two photon echo scheme $\theta_1 = 0.499\pi$ and $\theta_2 = 0.999\pi$. The echo strength of each echo reaches $\pi$ at a sufficiently large optical thickness, showing a possible enhancement of signal-to-noise ratio with tunable optical thickness. Fig.~\ref{fig:4}d) is the experimental measurement of two echoes. The inset schematic shows the pump pulses and the multi-echo train with a fixed time delay $\tau$. Fig.~\ref{fig:4}e) shows a similar result with three echoes. These observations of multi-echo pulse train demonstrate that the effective optically thick medium is achieved from an optically dilute material platform. High yield, low loss ($\sim$ 2.7 dB/m \cite{zhang2017monolithic}), meter-long waveguides in LN thin film are quickly becoming within reach thanks to the advancement of fabrication. Rich physics phenomena over the whole range of optical thickness are therefore easily accessible just in this single interface. In the analytical model by S. A. Moiseev, \textit{et al.} \cite{moiseev2020photon}, distinct and complex processes are involved in the third echo (Fig.~\ref{fig:4}f)). The first process is the self-stimulated echo from the two pumps and the second echo. The second process is a similar self-stimulated echo, involving the second pump pulse and the first two echoes. The third process represents the two photon echo from the first two echoes. The fourth process is a similar two photon echo from the first pump pulse and the first echo. The last process is the revival of the first echo, refocused by the second echo. With the unique waveguide geometry in the SmartCut thin film platform, we are able to extend our capability to experimentally probe this regime efficiently and effectively.

\section{Discussion}
In conclusion, we characterize SmartCut Er:LN thin film at cryogenic temperature down to 20 mK. The Smartcut technique allows integrating and miniaturizing optical devices with a direct light-ion interface, which is desirable in quantum processing and sensing technologies. Our sample devices incorporate Er$^{3+}$ ions into photonic nanostructures, which is telecommunication (telecomm.) wavelength compatible, allowing easy integration with the existing fiber optics technologies. This approach can also be easily extended to other rare earth elements, offering freedom to optimize for specific needs. We demonstrate that the SmartCut thin film is able to preserve bulk Er:LN coherence properties. In particular, the measured coherence time $T_\mathrm{m}$ can be as high as 180 $\mu$s, extending the knowledge of Er:LN optical coherence to the milli-Kelvin temperature regime. This long coherence time can be utilized to develop on-chip quantum memories at telecomm. wavelength \cite{saglamyurek2015quantum}, and is crucial for applications including spin qubit control \cite{zhou2017holonomic} and entanglement generation  \cite{bernien2013heralded}. By fabricating low-loss, long waveguides, we demonstrate the capability of optical thickness control, an added functionality favorable to many applications. By varying the waveguide lengths and excitation pulse strengths, we can control the number of echoes emitted, tailoring to applications of specific needs. A short waveguide can be used to generate just a single echo to ensure deterministic readout in a quantum memory. At the single photon level, this device has the potential to generate time bin entanglement of photons in two echo pulses. The intact bulk properties combined with the fabrication versatility in thin-film LN presents opportunities of complex devices for specific functions. Taking advantage of LN's excellent elctro-optic properties, dynamic control of optical emission \cite{casabone2021dynamic} can be realized by incorporating on-chip electrodes. The ability to integrate superconducting microwave structures  \cite{xu2021bedirectional} further opens up opportunities in microwave-optical interface  \cite{williamson2014magneto} and direct spin manipulation \cite{bienfait2016reaching}. 
\\

\section{Acknowledgement}
We thank Prof. Steven M. Girvin and Dr. Baptiste Royer for the thoughtful discussions. This work is supported by Department of Energy, Office of Basic Energy Sciences, Division of Materials Sciences and Engineering under Grant DE-SC0019406. The authors would like to thank Dr. Yong Sun, Sean Rinehart, Kelly Woods, and Dr. Michael Rooks for their assistance provided in the device fabrication. The fabrication of the devices was done at the Yale School of Engineering \& Applied Science (SEAS) Cleanroom and the Yale Institute for Nanoscience and Quantum Engineering (YINQE).\\

\noindent\textbf{REFERENCES}

\bibliography{References}

\end{document}